\documentclass[aps,prx,reprint,superscriptaddress]{revtex4-1}

\usepackage[dvips]{graphicx} 
\usepackage{amsfonts}
\usepackage{amssymb}
\usepackage{amscd}
\usepackage{amsmath}    
\usepackage{enumerate}
\usepackage{epsfig}
\usepackage{subfigure}
\usepackage{xcolor}
\usepackage{amsthm}
\usepackage{framed}

\newcommand{\bra}[1]{\mbox{$\left\langle #1 \right|$}}
\newcommand{\ket}[1]{\mbox{$\left| #1 \right\rangle$}}

\newcommand{\Pro}[1]{\mathrm{P(#1)}}

\definecolor{blue}{rgb}{0,0,1}
\definecolor{red}{rgb}{1,0,0}

\begin{document}
\title{Experimental Quantum Randomness Processing}
\date{\today}
\author{Xiao Yuan}
\author{Ke Liu}
\author{Yuan Xu}
\author{Weiting Wang}
\author{Yuwei Ma}
\author{Fang Zhang}
\author{Zhaopeng Yan}
\affiliation{Center for Quantum Information, Institute for Interdisciplinary Information Sciences, Tsinghua University, Beijing 100084, China}
\author{R.~Vijay}
\affiliation{Department of Condensed Matter Physics and Materials Science, Tata Institute of Fundamental Research, Homi Bhabha Road, Mumbai 400005, India}
\author{Luyan Sun}
\email{luyansun@tsinghua.edu.cn}
\author{Xiongfeng Ma}
\email{xma@tsinghua.edu.cn}
\affiliation{Center for Quantum Information, Institute for Interdisciplinary Information Sciences, Tsinghua University, Beijing 100084, China}

\begin{abstract}
Coherently manipulating multipartite quantum correlations leads to remarkable advantages in quantum
information processing. A fundamental question is whether such quantum advantages persist only by
exploiting multipartite correlations, such as entanglement. Recently, Dale, Jennings, and Rudolph negated
the question by showing that a randomness processing, quantum Bernoulli factory, using quantum
coherence, is strictly more powerful than the one with classical mechanics. In this Letter, focusing on the
same scenario, we propose a theoretical protocol that is classically impossible but can be implemented
solely using quantum coherence without entanglement. We demonstrate the protocol by exploiting the
high-fidelity quantum state preparation and measurement with a superconducting qubit in the circuit
quantum electrodynamics architecture and a nearly quantum-limited parametric amplifier. Our experiment
shows the advantage of using quantum coherence of a single qubit for information processing even when
multipartite correlation is not present.
\end{abstract}

\maketitle

Coherent superposition of different states, coherence, is a peculiar feature of quantum mechanics that distinguishes itself from Newtonian theory. In different scenarios, coherence exhibits as various quantum resources, such as entanglement \cite{Horodecki09}, discord \cite{Modi12}, and single-party coherence \cite{Baumgratz14}. In many quantum information tasks, the common resource leading to quantum advantage is multipartite quantum correlations. For instance, entanglement plays a crucial role in quantum key distribution \cite{bb84,Ekert1991}, teleportation \cite{Teleport1993}, and computation \cite{Shor97,Grover96}. While the essence of multipartite correlation originates from coherent superposition, it is natural to expect the essence of quantum advantage to also originate from coherence. This raises a fundamental question: Can quantum advantage be obtained without using multipartite correlations?

\begin{figure}[bht]
\centering
\resizebox{8cm}{!}{\includegraphics[scale=1]{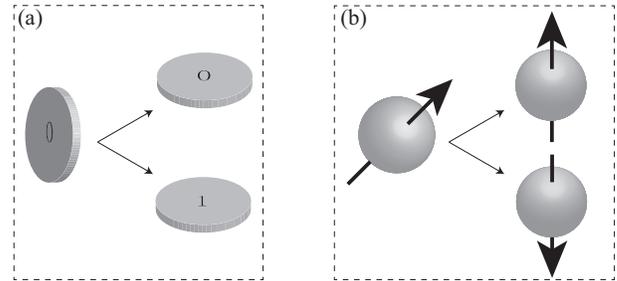}}
\caption{Classical and quantum coin. For a given $p$ value, (a) classical and (b) quantum $p$-coin corresponds to two different ways of encoding $p$, see Eqs.~\eqref{Eq:classicalcoin} and \eqref{Eq:quantumcoin}, respectively. The key difference lies in whether there is coherence in the computational basis.}\label{fig:coin}
\end{figure}

In randomness generation, it has been shown that coherence is the essential resource for generating true random numbers \cite{Yuan15}. It is thus natural to expect coherence to be a
resource for displaying quantum advantages in certain randomness related tasks. Remarkably, in a recent work by Dale \emph{et al.}, a rather simple task of randomness processing is proposed to show that coherence yields a provable quantum advantage over classical stochastic physics \cite{dale2015provable}. In this randomness processing task, a classical coin, see Fig.~\ref{fig:coin}(a), corresponds to a classical machine that produces independent and identically distributed random variables where each one has the binary values, head (0) and tail (1). A coin is called $p$-coin if the probability of producing a head is $p$, where $p\in[0,1]$. Given an unknown $p$-coin, an interesting question is whether one can construct an $f(p)$-coin, where $f(p)$ is a function of $p$ and $f(p)\in[0,1]$. Such construction processing is called a Bernoulli factory \cite{Asmussen92, RSA:RSA20333}.

Let us take $f(p)=1/2$ for example, which was solved by von Neumann with a rather simple but heuristic strategy \cite{von195113}. Flip the $p$-coin ($p\ne0$) twice. If the outcomes are the same start over; otherwise, output the second coin value as the $1/2$-coin output. Therefore, the function of $f(p)=1/2$ can be constructed from an arbitrary unknown $p$-coin. As a generalization, a natural question involves which kind of function $f(p)$ can be constructed from an unknown $p$-coin. This classical Bernoulli factory problem was solved by Keane and O'Brien \cite{Keane94}. Generally speaking, a necessary condition for $f(p)$ being constructible is that $f(p)\neq 0$ or $1$ when $p\in(0,1)$. The function $f(p)=1/2$ satisfies this condition, while there are many other examples that violate it. For instance, surprisingly, the simple ``probability amplification'' function $f(p)=2p$ \footnote{A complete definition is: $f(p)=2p$ when $p\in[0,1/2]$ and $f(p)=2(1-p)$ when $p\in(1/2,1]$.} does not satisfy the constructible condition, where we have $f(1/2)=1$. Therefore, there is no classical method to construct an $f(p)=2p$-coin.

In the language of quantum mechanics, a $p$-coin corresponds to a machine that outputs identically mixed qubit states,
\begin{equation}\label{Eq:classicalcoin}
  \rho_C = p\ket{0}\bra{0} + (1-p)\ket{1}\bra{1},
\end{equation}
where $p\in[0,1]$, and $Z = \{\ket{0},\ket{1}\}$ is the computational basis denoting head and tail, respectively. As $p$ is generally unknown, we can regard $\rho_C$ as a classical way of encoding an unknown parameter $p$. A measurement in the $\{\ket{0},\ket{1}\}$ basis would output a head or a tail with a probability according to $p$ and $1-p$, respectively. On the other hand, a quantum way of encoding $p$, see Fig.~\ref{fig:coin}(b), can be a coherent superposition of $\ket{0}$ and $\ket{1}$, i.e., $\rho_Q = \ket{p}\bra{p}$ with
\begin{equation}\label{Eq:quantumcoin}
  \ket{p} = \sqrt{p}\ket{0}+\sqrt{1-p}\ket{1}.
\end{equation}
Following the nomenclature in Ref.~\cite{dale2015provable}, we call such a quantum coin a \emph{quoin}. It is straightforward to see that a $p$-coin can always be constructed from a $p$-quoin by measuring it in the $Z$ (computational) basis. Thus, classically constructible (via coins) $f(p)$ functions are also quantum mechanically constructible (via quoins), while a really interesting question is whether the set of quantum constructible functions (via quantum a Bernoulli factory) is strictly larger than the classical set.


In Ref.~\cite{dale2015provable}, Dale \emph{et al.} have theoretically proved the necessary and sufficient conditions for $f(p)$ being quantum constructible. Specifically, they show that there are functions, for instance $f(p)=2p$, which are impossible to construct classically, but can be efficiently realized in the presence of $p$-quoins. Therefore, they provide a positive answer to this problem where quantum resources are strictly superior to classical ones. The protocol for generating the $f(p)=2p$ function relies on Bell state measurement on two quoins, which essentially establish entanglement between the two quoins.

Now, we are interested in seeing whether such a quantum advantage persists even when multipartite correlations, such as entanglement, are absent. Thus, we only allow single qubit operations. Without two-qubit operations, it turns out that constructing the $f(p)=2p$ function will require many copies of qubits defined in Eq.~\eqref{Eq:quantumcoin} and the convergence could be poor. In this Letter, we propose another function that is impossible with classical means but feasible with only limited number of single-qubit operations.


\emph{Theoretical protocol} --- In this Letter, we analyze a classically impossible $f(p)$ function defined by
\begin{equation}\label{Eq:fp}
  f(p) = 4p(1-p).
\end{equation}
For $p=1/2$, we have $f(p)=1$ which means that this function is classically unachievable. On the other hand, it is straightforward to check that the $f(p)$-function satisfies the requirements for beign quantum constructible \cite{dale2015provable}. Given a $p$-quoin, we explicitly present an efficient protocol for generating an $f(p)$-coin as shown in Table~\ref{Table:protocol}.

\begin{table}[htbp]
\caption{A protocol for generating an $f(p)=4p(1-p)$-coin from a $p$-quoin.} \label{Table:protocol}
\begin{framed}
\centering
\begin{enumerate}[Step 1]

\item
Generate a $p$-coin: measure a quoin, $\ket{p}$ defined in Eq.~\eqref{Eq:quantumcoin}, in the $Z$ basis. The measurement outcome is a $p$-coin.

\item
Generate a $q$-coin, where $q = \left[1+2\sqrt{p(1-p)}\right]/2$: measure the same quoin $\ket{p}$  in the $X=\{(\ket{0}+\ket{1})/\sqrt{2},(\ket{0}-\ket{1})/\sqrt{2}\}$ basis. The measurement outcome is a $q$-coin.

\item
Construct an $m$-coin from a $p$-coin, where $m=2p(1-p)$: toss the $p$-coin twice, output head if the two tosses are different and tail otherwise. Similarly, one can construct an $n$-coin from a $q$-coin, where $n=2q(1-q)=1/2-2p(1-p)$.

\item
Construct an $s$-coin from an $m$-coin, where $s=m/(m+1)$: toss the $m$-coin twice, if the first toss is tail then output tail; otherwise if the second toss is tail, output head; otherwise if both tosses are head, repeat this step. Similarly, one can construct a $t$-coin from an $n$-coin, where $t=n/(n+1)$.

\item
Construct an $f(p)=4p(1-p)$-coin: first toss the $s$-coin and then the $t$-coin. If the first toss is head and the second toss is tail, then output head; if the first toss is tail and the second toss is head, then output tail; otherwise repeat this step.

\end{enumerate}
\end{framed}
\end{table}

In our protocol, generating the $q$-coin, where
\begin{equation}\label{eq:qcoins}
   q= \frac{1}{2}\left[1+2\sqrt{p(1-p)}\right],
\end{equation}
is an essential nonclassical step. In fact, the only additionally required coin for constructing all quantum constructible $f(p)$-coins is the $h_a(p)$-coin,
\begin{equation}\label{}
  h_a(p) = \left(\sqrt{p(1-a)} + \sqrt{a(1-p)}\right)^2,
\end{equation}
which can be obtained by measuring the quoin in the $\{\sqrt{1-a}\ket{0}+\sqrt{a}\ket{1}, \sqrt{a}\ket{0}-\sqrt{1-a}\ket{1}\}$ basis. In our case, we set $a=1/2$. Here, one can see that entanglement is not necessary to quantum Bernoulli factory.

As shown in Table \ref{Table:protocol}, the first two steps involve quantum devices, where quoins are measured in the $Z$ and $X=\{(\ket{0}+\ket{1})/\sqrt{2},(\ket{0}-\ket{1})/\sqrt{2}\}$ bases, respectively, to obtain the $p$- and $q$-coins. The following steps (step 3-5) are classical processing of the $p$- and $q$-coins. The rigourous derivation of the classical steps can be found in the Appendix. Comparing to the $f(p)=2p$ function, our protocol converges much faster, which results in a higher fidelity for the realization.

In practice, owing to experimental imperfections, we cannot realize perfect $p$-quoins and perform ideal measurements to get perfect $p$- and $q$-coins. Thus, in reality, we cannot realize exact $f(p)$-coins, especially, we cannot get $f(p) = 1$ when $p=1/2$. Following previous studies \cite{nacu2005fast, Mossel05, thomas2011practical}, we employ a truncated function
\begin{equation}\label{Eq:truncated}
  f_{t} = \min \{f, 1-\epsilon\},
\end{equation}
with $\epsilon$ describing the imperfections. When $\epsilon$ is nonzero, the truncated function of  $f=4p(1-p)$ falls in the classical Bernoulli factory and hence can be constructed via $p$-coins. However, the number of classical coins $N$ required to construct $f(p)$ scales poorly with $\epsilon$, see Appendix for more details. In the experiment, we need to implement high fidelity state preparation and measurement to reduce $\epsilon$ as small as possible in order to faithfully demonstrate the quantum advantage.

In the following, we focus on the preparation and measurement of the $p$-quoin, and how to construct an $f(p)=4p(1-p)$ coin via necessary classical processing. Here, we emphasize that the quantum circuit to realize the operations should be independent of $p$. In demonstration, we fix the measurement setting and prepare $p$-quoins for various $p$ values.


\emph{Experimental realization} --- We choose a superconducting qubit system to prepare $p$-quoins. Superconducting quantum systems have made tremendous progress in the last decade, including realizing long coherence times, showing great stability with fast and precise qubit manipulations, and demonstrating high fidelity quantum non-demolition (QND) qubit measurement. Thus, it makes a perfect candidate for our test.

\begin{figure}[b]
\centering
\resizebox{8cm}{!}{\includegraphics[scale=1]{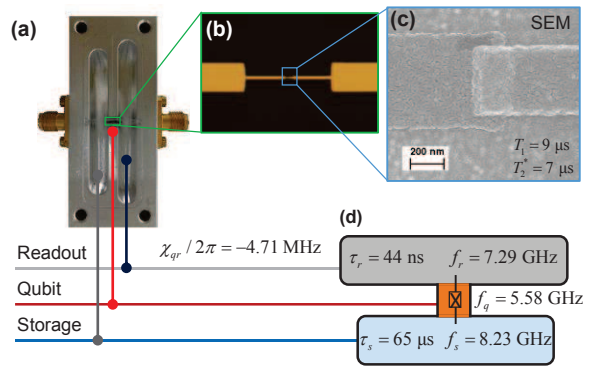}}
\caption{Experimental setup. (a) Optical image of a transmon qubit located in a trench, which dispersively couples to two 3D Al cavities. (b) Optical image of the single-junction transmon qubit. (c) Scanning electron microscope image of the Josephson junction. (d) Schematic of the device with the main parameters. In our experiment, the higher frequency cavity is not used and always remains in vacuum, which can be used as another $p$-quoin in future experiments \cite{Leghtas2}. Note that the highlighted boxes in (a) and (b) are not to scale and are intended for illustrative purposes only.}
\label{fig:device}
\end{figure}

In our experiment, we employ the so-called `circuit quantum electrodynamics architecture' \cite{Wallraff}. A superconducting transmon qubit (our quoin) is located in a waveguide trench and dispersively couples to two 3D cavities~\cite{Kirchmair,Vlastakis,SunNature} as shown in Fig.~\ref{fig:device}. The transmon qubit has a transition frequency of $\omega_q/2\pi=5.577$ GHz, an anharmonicity $\alpha_q/2\pi=-246$ MHz, an energy relaxation time $T_1=9~\mu$s, and a Ramsey time $T_2^*=7~\mu$s. The larger cavity has a resonant frequency of $\omega_c/2\pi=7.292$ GHz and a decay rate of $\kappa/2\pi=3.62$ MHz, which provides a fast way of reading out the qubit state through their strong dispersive interaction with a dispersive shift $\chi/2\pi=-4.71$~MHz. As we focus on exhibiting quantum advantage solely with a single quantum system, the smaller cavity with a higher resonant frequency is not used and remains in a vacuum state. This higher frequency cavity can potentially be used as another $p$-quoin in future experiments \cite{Leghtas2}. In this case, joint measurement can be performed on two $p$-quoins, which may save the resource. For now, we focus on single-qubit operations.

The output of the readout cavity is connected to a Josephson parametric amplifier (JPA)~\cite{Hatridge,Roy}, operating in a double-pumped mode~\cite{Kamal,Murch} as the first stage of amplification between the readout cavity, at a base temperature of 10~mK, and the high electron mobility transistor, at 4~K. To minimize pump leakage into the readout cavity and achieve a longer $T_2^*$ dephasing time, we operate the JPA in a pulsed mode. The readout pulse width has been optimized to 180~ns with a few photons in order to have a high signal-to-noise ratio. This JPA allows a high-fidelity single-shot readout of the qubit state. The overall readout fidelity of the qubit measured for the ground state $\ket{0}$ when initially prepared at $\ket{0}$ by a post-selection is 0.996, demonstrating the high QND nature of the readout, while the fidelity for the excited state $\ket{1}$ is slightly lower, 0.943 (see Appendix). The loss of both fidelities is predominantly limited due to the $T_1$ process during both the waiting time of the initialization measurement (300~ns) and the qubit readout time (180~ns).

Due to stray infrared photons and other background noise, our qubit has an excited state population of about $8.5\%$ in the steady state. The high QND qubit measurement allows us to eliminate these imperfections by performing an initialization measurement to purify the qubit by only selecting the ground state for the following experiments \cite{Riste}. The measurement pulse sequences for preparing quoins can be found in the Appendix. It is worth mentioning that our superconducting system always yields a detection result once the measurement is performed, which is very challenging for other implementations, such as lossy photonic systems.

We apply an on-resonant microwave pulse to rotate the qubit to an arbitrary angle $\theta$ along the $Y$-axis, $R_\theta^Y = \exp(-i\sigma_y\theta/2)$, where $\sigma_y$ is the Pauli matrix, for a preparation of any $p = \cos^2(\theta/2)$-quoins. We  choose a gaussian envelope pulse truncated to $4\sigma=24$~ns for the rotation operations. We also use the so-called ``derivative removal by adiabatic gate" \cite{Motzoi} technique to minimize qubit leakage to higher levels outside the computational space. A randomized benchmark calibration~\cite{Knill2008,Ryan2009,Magesan2012,BarendsNature} shows that the $R_{\pi/2}^{Y}$ gate fidelity itself is about 0.998, mainly limited by the qubit decoherence (see Appendix). The final measurement for the quoins is along either the $Z$-axis or the $X$-axis. The measurement along the $X$-axis is realized by applying an extra $R_{\pi/2}^{-Y}$ rotation (Hadamard transformation) followed by a $Z$-basis measurement.

\begin{figure}[bht]
\centering
{\resizebox{8cm}{!}{\includegraphics[scale=1]{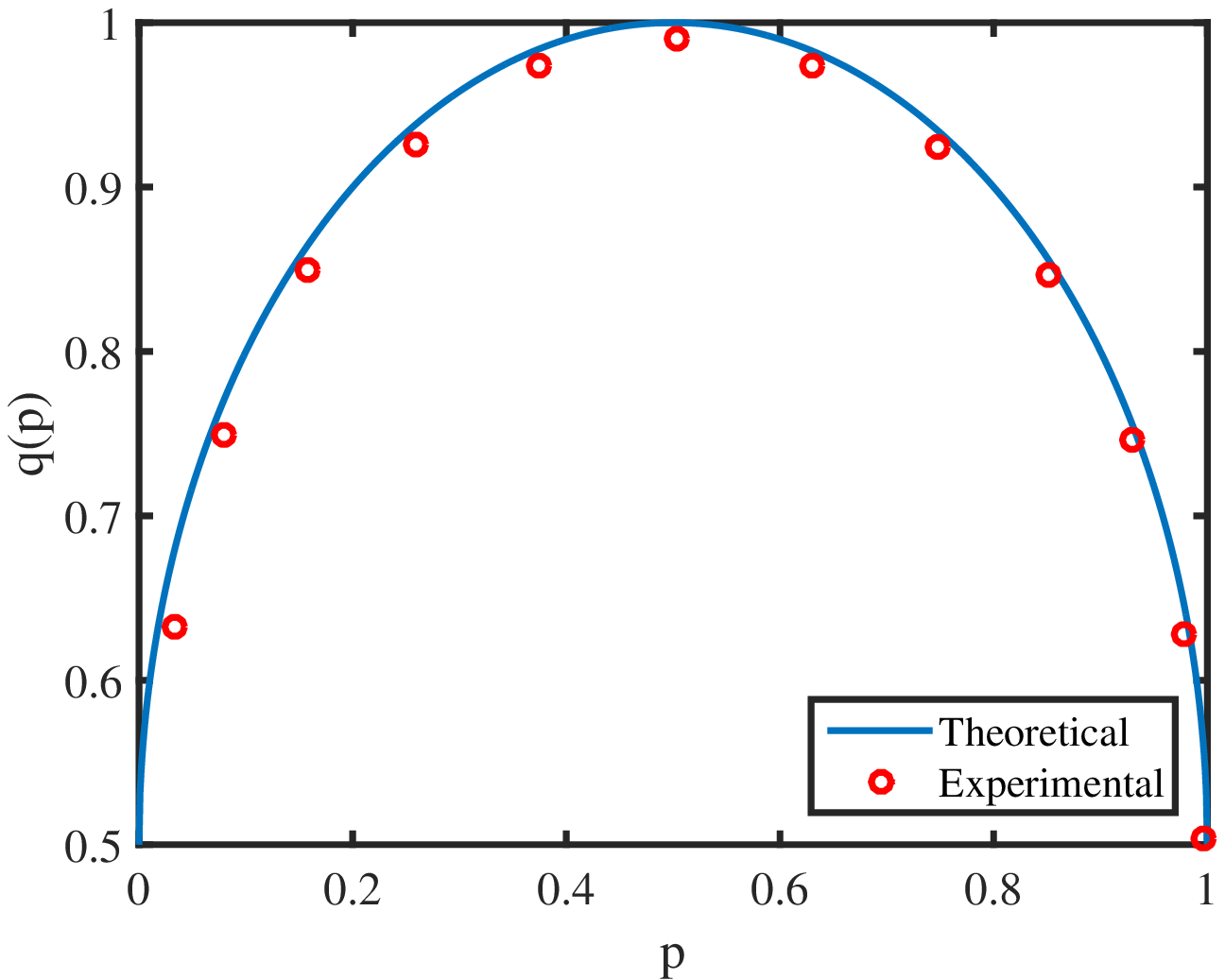}}}
{\resizebox{8cm}{!}{\includegraphics[scale=1]{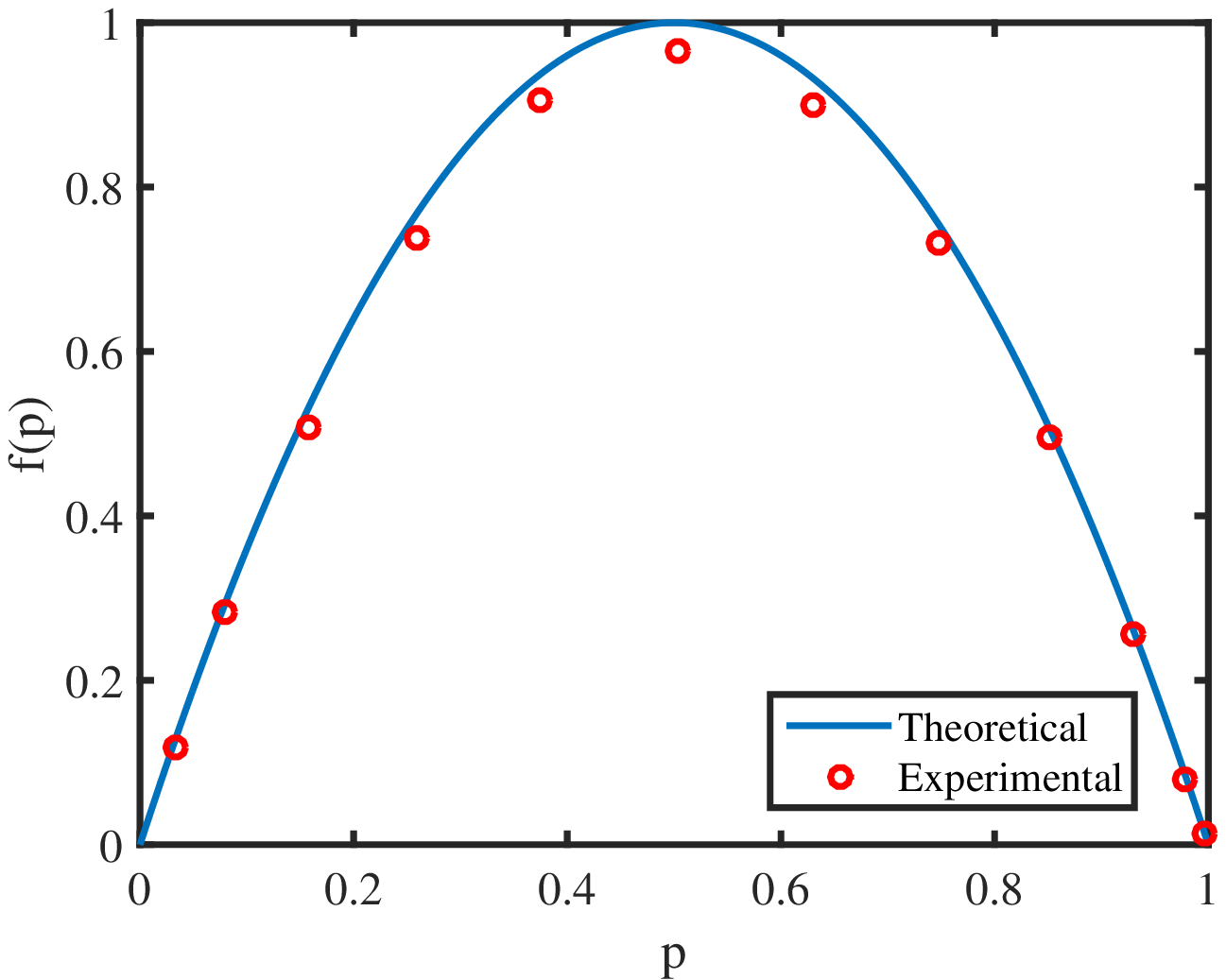}}}
\caption{Theoretical and experimental results for the (a) $q$-coin and (b) $f(p)=4p(1-p)$-coin. 
Here, the number of experiment data for the $p$-quoins is in the order of $10^7$ and the number for the $f(p)$-coin is in the order of $10^6$. On average, we need about 20 $p$-quoins to construct a $f(p)$-coin. The standard deviations of $p$, $q$, and $f(p)$ are in the order of $10^{-4}$, thus are not plotted in the figure.}\label{fig:result}
\end{figure}

In our experiment, the $q$-coins as defined in Eq.~\eqref{eq:qcoins} are implemented, which are also classically impossible \footnote{The function $q(p)$ is defined in Eq.~\eqref{eq:qcoins}. It is straightforward to check to see that $q(1/2)=1$, indicating that it is classically impossible} when regarded as a function of $q$. We plot the experiment result of the $q$-coins in Fig.~\ref{fig:result}(a) and the result of the $f(p) = 4p(1-p)$-coins by following the protocol in Table~\ref{Table:protocol} in Fig.~\ref{fig:result}(b). The experimentally realized values of $q_{\mathrm{exp}}$ and $f_{\mathrm{exp}}(p)$ are sampled from the observed coins, which match well with the theoretical predictions. By implementing state preparation, operation and measurement with high fidelities, we are able to achieve $q_{\mathrm{exp}}(1/2) = 0.990$ and $f_{\mathrm{exp}}(1/2)=0.965$, which can be well modeled by the truncated function defined in Eq.~\eqref{Eq:truncated} with $\epsilon = 0.010$ and $\epsilon = 0.035$, respectively.

\emph{Discussion} ---
The classical Bernoulli factory cannot produce exact $q$- and $f(p) = 4p(1-p)$-coins with finite number of usages of $p$-coins. In practice, the implemented function may deviate from the desired one due to device imperfections. In this case, the practically realized coins may be constructible with classical means, though the number of classical coins required may increase drastically with decreasing deviation. Focusing on the truncated function defined in Eq.~\eqref{Eq:truncated}, we present a classical protocol for simulating the experiment data $f_{\mathrm{exp}}(p)$ with $\epsilon = 0.035$ in the Appendix. It is shown that, on average, more than $10^4$ classical $p$-coins are required for constructing the truncated function, which is much larger than the average number of quoins (about $20$) used in our protocol \footnote{Strictly speaking, the quantum advantage demonstrated here is a weaker version of the one mentioned in the beginning of the Letter, where the function is classically impossible to construct.}. For the $q$-coin, as the deviation is smaller, the classical simulation is even harder. In the Appendix, we show that more than $10^5$ classical coins are needed for the truncated function, while our quantum protocol only requires one quoin.

From the experimental perspective, the small deviation $f_{exp}(1/2)$ from unity in the ideal case is dominated by qubit decoherence. With better qubit coherence times of $T_1, T_2\sim 100~\mu$s achieved recently~\cite{Rigetti}, we expect the deviation of $f_{exp}(p)$ from $f_{th}(p)$ to be an order of magnitude lower. In future, a more accurate quantum Bernoulli factory can be realized and the classical simulation will eventually become intractable.


In quantum Bernoulli factory, the only resource that is responsible for constructing a classically impossible function is the quantumness of single qubits --- coherence. Our experiment also involves only single-qubit operations and hence proves the quantum advantage solely using coherence without multipartite correlations. Recently, a coherence framework \cite{Baumgratz14} is proposed in which coherence can be measured quantitatively. Along this line, it would be interesting to see whether the advantage of constructing $f(p)$ from $p$-quoins is directly related to the coherence of the $p$-quoins. Note that Bernoulli factory is a randomness process. From the close relation between randomness and coherence \cite{Yuan15}, we expect that a general $p$-quoin with larger coherence would show advantage over $p$-quoin with smaller coherence.

Our experiment verification sheds light on a fundamental question about what is the essential resource for quantum information processing, which may stimulate the search for more protocols that show quantum advantages without multipartite correlations. Considering the conversion from coherence to multipartite correlation \cite{Streltsov15}, investigating the power of coherence may also be helpful for understanding the power of multipartite correlation and universal quantum computation \cite{howard2014contextuality}.

It is noteworthy that entanglement can be exploited to save resource in the quantum Bernoulli factory, which provides an extra advantage for randomness processing \cite{dale2015provable}. Extending our implementation to multi-qubit systems can verify this extra quantum advantage. When considering practical imperfections, multiple qubit operation generally has a lower fidelity of measurement. Balancing between the saving of resource and decoherence due to multiple-qubit interactions, it is interesting to see whether multipartite correlation can have extra advantage in practice. As we are focusing on proving the advantage only with coherence, we leave such extension and discussion to future works.


We acknowledge Z.~Cao, H.~Dale, E.~Ginossar, D.~Jennings, T.~Rudolph, D.~Schuster, and H.~Zhou for the insightful discussions and thank M.~Chand and T.~Roy for the help with parametric amplifier measurements. This work was supported by the National Basic Research Program of China Grants No.~2011CBA00300 and No.~2011CBA00301, the 1000 Youth Fellowship program in China, and by the National Natural Science Foundation of China Grants No.~11474177.

X.Y.~and K.L.~contributed equally to this work.

\appendix

\section{Protocol for $f(p)=4p(1-p)$}
A (classical) coin corresponds to a classical machine that produces independent and identically distributed (i.i.d.) random variables where each random variable has binary values, head (0) and tail (1). A coin is called $p$-coin, if the probability of producing head is $p$, where $p\in[0,1]$. In quantum mechanics, a $p$-coin corresponds to a machine that outputs identically mixed qubit states,
\begin{equation}\label{Eq:classicalcoin}
  \rho_C = p\ket{0}\bra{0} + (1-p)\ket{1}\bra{1},
\end{equation}
where $p\in[0,1]$, and $Z = \{\ket{0},\ket{1}\}$ is the computational basis corresponding to head and tail, respectively. On the other hand, a quantum way of encoding $p$ (quoin) can be a coherent superposition of $\ket{0}$ and $\ket{1}$, i.e., $\rho_Q = \ket{p}\bra{p}$ with
\begin{equation}\label{Eq:quantumcoin}
  \ket{p} = \sqrt{p}\ket{0}+\sqrt{1-p}\ket{1}.
\end{equation}

The protocol of generating a $f(p) = 4p(1-p)$-coin with a $p$-quoin is shown in Table I of Main Text. Now, we analyze each step in detail.
\begin{enumerate}[Step 1]
\item
\textbf{Generate a $p$-coin}:

When measuring a $p$-quoin, as given by Eq.~\eqref{Eq:quantumcoin}, in the $Z$ basis, the probabilities of obtaining $0$ and $1$ are $p$ and $1-p$, respectively.

\item
\textbf{Generate a $q$-coin}, where $q = \left[1+2\sqrt{p(1-p)}\right]/2$:

When measuring a $p$-quoin in the $X=\{(\ket{0}+\ket{1})/\sqrt{2},(\ket{0}-\ket{1})/\sqrt{2}\}$ basis, the probabilities of obtaining $(\ket{0}+\ket{1})/\sqrt{2}$ and $(\ket{0}-\ket{1})/\sqrt{2}$ are $\left[1+2\sqrt{p(1-p)}\right]/2$ and $\left[1-2\sqrt{p(1-p)}\right]/2$, respectively.

\item
\textbf{Construct an $m$-coin from a $p$-coin}, where $m=2p(1-p)$: toss the $p$-coin twice, output head if the two tosses are different and tail otherwise.

The probability of output two different tossing result is
\begin{equation}\label{}
m=\Pro{head}\Pro{tail} + \Pro{tail}\Pro{head} = 2p(1-p).
\end{equation}
Similarly, one can construct an $n$-coin from a $q$-coin, where $n=2q(1-q)=1/2-2p(1-p)$.

\item
\textbf{Construct an $s$-coin from an $m$-coin, where $s=m/(m+1)$}: toss the $m$-coin twice, if the first toss is tail then output tail; otherwise if the second toss is tail, output head; otherwise, repeat this step.

Denote the probability of outputting head and tail by $\Pro{H}$ and $\Pro{T}$, respectively, then,
\begin{equation}\label{}
\begin{aligned}
\Pro{H} &= \Pro{head}\Pro{tail} + \Pro{head}(1-\Pro{tail})\Pro{H}\\
& = m(1-m) + m^2\Pro{H}.
\end{aligned}
\end{equation}
Solving this equation, we have
\begin{equation}\label{}
s=\Pro{H} = \frac{m}{m+1}.
\end{equation}
Similarly, one can construct a $t$-coin from an $n$-coin, where $t=n/(n+1)$.

\item
\textbf{Construct an $f(p)=4p(1-p)$-coin}: first toss the $s$-coin and then the $t$-coin. If the first toss is head and the second toss is tail, then output head; if the first toss is tail and the second toss is head, then output tail; otherwise repeat this step.

Denote the probability of outputting head and tail by $\Pro{H}$ and $\Pro{T}$, respectively, then,
\begin{equation}\label{}
\begin{aligned}
  \Pro{H} =& \Pro{s-head}\Pro{t-tail} + (1 - \Pro{s-head}\Pro{t-tail}
  \\& - \Pro{s-tail}\Pro{t-head})\Pro{H}\\
  = &s(1-t) + \left[1 - s(1-t) - (1-s)t\right]\Pro{H}\\
  = &\frac{m}{m+1}\left(1-\frac{n}{n+1}\right)  \\
  &+\Pro{H}\left[1 - \frac{m}{m+1}\left(1-\frac{n}{n+1}\right) - \left(1-\frac{m}{m+1}\right)\frac{n}{n+1}\right]\\
    = &\frac{m}{(m+1)(n+1)} + \frac{mn + 1}{(m+1)(n+1)} \Pro{H}
\end{aligned}
\end{equation}
Solving this equation, we have
\begin{equation}\label{}
\begin{aligned}
  \Pro{H} &= \frac{m}{m+n} \\
  & = \frac{2p(1-p)}{2p(1-p)+1/2-2p(1-p)}\\
  &= 4p(1-p).
\end{aligned}
\end{equation}
\end{enumerate}

Therefore, we just prove our protocol of constructing the $f(p) = 4p(1-p)$ function.

\section{Simulation of Experiment data}
\subsection{The truncated function}\label{Sec:trun}
Here, we show how to construct the truncated function
\begin{equation}\label{Eq:truncated}
  f_t(p) = \min\{4p(1-p), 1-\epsilon_1\}, \epsilon_1=0.035
\end{equation}
from $p$-coin by classical means. The protocol works as follows.
\begin{enumerate}[(i)]
\item
Toss the $p$-coin twice, if the outputs are different, then output head otherwise output tail. This achieves $g(p)=2p(1-p)$-coin with two $p$-coins.
\item
Apply Theorem 1 in Ref.~\cite{nacu2005fast}, which gives $h(p)= \min\{2p, 1- 2\epsilon_1'\}$, and perform the composition $h(g(p))=\min\{4p(1-p), 1- 2\epsilon_1'\}$. Let $\epsilon_1'= 0.0175$, the desired function is obtained.
\end{enumerate}
Now, we calculate the number of $p$-coins needed in step (ii). By Theorem 1 in Ref.~\cite{nacu2005fast}, the probability that more than $n$ $p$-coins are needed is bounded by
\begin{equation}\label{Eq:pn}
 \begin{aligned}
 P(N>n) \le & \frac{\sqrt{2}}{\epsilon_1' (\sqrt{2}-1)}\sqrt{\frac{2}{n}} e^{-2\epsilon_1'^2 n}\\
 & + 72(1- e^{-2 \epsilon_1'^2})^{-1} e^{-2\epsilon_1'^2 n}+ 4\cdot  2^{-n/9}.
 \end{aligned}
\end{equation}
With large $n$ and small $\epsilon_1'=0.0175$, we can approximate Eq.~\eqref{Eq:pn} by
\begin{equation}\label{Eq:}
\begin{aligned}
P(N>n) \lessapprox &\frac{4.8}{\epsilon_1' \sqrt{n}}e^{-2\epsilon_1'^2 n} + \frac{36}{\epsilon_1'^2}e^{-2\epsilon_1'^2 n}\\
\approx & \frac{36}{\epsilon_1'^2}e^{-2\epsilon_1'^2 n}.
\end{aligned}
\end{equation}
This bound is nontrivial only if the right hand side is less and equal to 1, that is,
\begin{equation}\label{}
  n \approx \frac{-1}{2\epsilon_1'^2}\ln\left(\frac{\epsilon_1'^2}{36}\right) \approx 1.9 \times 10^4.
\end{equation}

Thus combining (i) and (ii), the number of $p$-coins needed to simulate the $f_t(p)$ function is more than $2\times 1.9\times 10^4= 3.8\times 10^4$. Note that, Eq.~\eqref{Eq:pn} provides only an upper bound to the probability distribution, there may exists more efficient protocols that requires less number usages of $p$-coins. 

%

\subsection{Simulation of the $q$-coin}
Here, we show how to simulate the truncated function of $q$ coin,
\begin{equation}\label{}
  q_t(p)=\min  \left\{\frac{1}{2}\left[1+2\sqrt{p(1-p)}\right], 1-\epsilon_3\right\},
\end{equation}
with $\epsilon_3=0.01$. To do so, we first construct the truncated coin $f_t(p)$ defined in Eq.~\eqref{Eq:truncated}.  Then we can simulate $q_t(p)$ with the $f_t(p)$-coin by applying the following protocol,
\begin{enumerate}
  \item Apply a square root function of $f_t(p)$, which gives a $\sqrt{f_t(p)}$-coin.
  \item Toss the 1/2-coin and the $\sqrt{f_t(p)}$-coin, output tail if both tosses are tail.
\end{enumerate}
Then, it is straightforward to check that the following coin is prepared
\begin{equation}\label{}
\begin{aligned}
  Q_t(p)&=\frac{1}{2}\left[1+\sqrt{f_t(p)}\right]\\
  & = \min  \left\{\frac{1}{2}\left[1+2\sqrt{p(1-p)}\right], \frac{1}{2}\left[1+\sqrt{1-\epsilon_1}\right]\right\},
\end{aligned}
\end{equation}
which coincides with the $q_t(p)$-coin if we let
\begin{equation}\label{}
  1-\epsilon_3 = \frac{1}{2}\left[1+\sqrt{1-\epsilon_1}\right].
\end{equation}
In this case, we have $\epsilon_1 = 0.04$. To simulate the $f_t(p)$-coin, we can follow the protocol in Sec.~\ref{Sec:trun}, which costs more than $4\times10^4$ number of $p$-coins on average for each $f_t(p)$-coin. The square root function of $f_t(p)$ can be constructed by following the method from Ref.~\cite{Mossel05} or the one presented in Ref.~\cite{dale2015provable}. On average, more than $10$ coins are needed for constructing the square root function. Therefore, more than $4\times10^5$ number of $p$-coins are necessary for the construction of the truncated function $q_t(p)$.

\section{Experiment setup and results}
\begin{figure}[bht]
\centering
\resizebox{8cm}{!}{\includegraphics[scale=1]{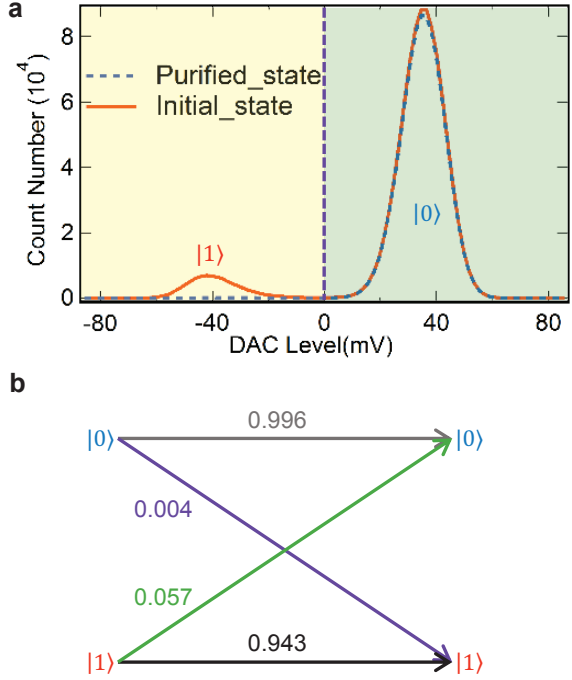}}
\caption{Readout properties of the qubit. The phase between the JPA readout signal and the pump has been adjusted such that $\ket{0}$ and $\ket{1}$ states can be distinguished with optimal contrast. a) Bimodal and well-separated histogram of the qubit readout. A threshold $V_{th}=0$ has been chosen to digitize the readout signal. Solid line is for an initial measurement showing about 8.5\% $\ket{1}$ state, while dashed line is for a second measurement after initially selecting $\ket{0}$ state. The disappearance of $\ket{1}$ state demonstrates both a high purification and high quantum non-demolition measurement of the qubit. b) Basic qubit readout matrix. The loss of fidelity predominantly comes from the $T_1$ process during both the waiting time after the initialization measurement and the qubit readout time.}
\label{fig:ReadoutProperty}
\end{figure}

The readout property of the qubit is first characterized as shown in Fig.~\ref{fig:ReadoutProperty}. The smaller cavity has a resonant frequency of $\omega_s/2\pi=8.229$ GHz and remains in vacuum all the time. Because we always purify our qubit initial state to the ground state $\ket{0}$ and use pulses with DRAG~\cite{Motzoi} to minimize the leakage to levels higher than the first excited state $\ket{1}$, we do not distinguish the levels higher than $\ket{1}$ in the readout. We thus adjust the phase between the JPA readout signal and the pump such that $\ket{0}$ and $\ket{1}$ states can be distinguished with optimal contrast. Figure~\ref{fig:ReadoutProperty}a shows the histogram of the qubit readout. The histogram is clearly bimodal and well-separated. A threshold $V_{th}=0$ is chosen to digitize the readout signal.

\begin{figure}[bht]
\centering
\resizebox{8cm}{!}{\includegraphics[scale=1]{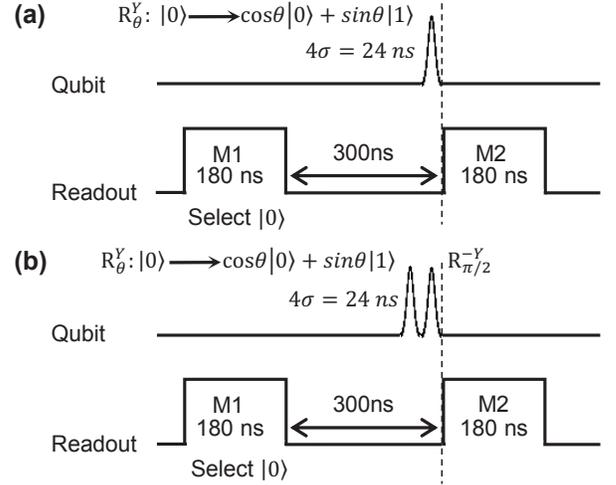}}
\caption{Experimental pulse sequences for the preparation of quoins and the measurements in the $Z$ (a) and $X$ (b) bases. An initial measurement M1 is firstly performed to purify the qubit to the ground state $\ket{0}$. The rotation of the qubit is realized by applying an on-resonance microwave pulse with various amplitudes. The measurement is always performed in the $Z$-basis. The measurement in the $X$-basis is realized by performing an extra $R_{\pi/2}^{-Y}$ pre-rotation. The phase of this extra pre-rotation is chosen to minimize the effect from qubit decoherence during the measurement for the case of $p=0.5$, which is most sensitive to the final qubit readout accuracy.}
\label{fig:quantum circuit}
\end{figure}

Due to stray infrared photons or other background noises, our qubit has an excited state population of about $8.5\%$ in the steady state (solid histogram in Fig.~\ref{fig:ReadoutProperty}a). In order to eliminate these excited states for the quoin experiments, a high quantum non-demolition qubit measurement M1 is performed to allow a qubit purification by only selecting $\ket{0}$ state (see Fig.~\ref{fig:quantum circuit})~\cite{Riste}. We wait 300~ns for the readout photons to leak out before the preparation of the qubit to arbitrary superposition states through an on-resonant microwave pulse with various amplitudes. After a purification to $\ket{0}$ state, the following measurement gives a probability of 0.996 of $\ket{0}$ state (dashed histogram in Fig.~\ref{fig:ReadoutProperty}a), demonstrating the high quantum non-demolition nature of the qubit measurement. Figure~\ref{fig:ReadoutProperty}b shows the basic qubit readout properties. The readout fidelity of the qubit measured at $\ket{1}$ state while initially prepared at $\ket{1}$ state by a measurement is 0.943. The loss of both fidelities is predominantly limited due to the $T_1$ process during both the waiting time of the initialization measurement (300~ns) and the qubit readout time (180~ns).

The experimental pulse sequences for the quoins with state preparations are shown in Fig.~\ref{fig:quantum circuit}. The measurement is always performed in the $Z$ basis. The $X$-basis measurement is realized by performing an extra $R_{\pi/2}^{-Y}$ rotation before the Z-basis measurement. The phase of this extra pre-rotation is chosen to minimize the effect from qubit decoherence during the measurement. In experiment, we prepare the qubits to $\ket{\psi} = \cos(\theta/2)\ket{0} +(\sin\theta/2)\ket{1}$. For different $\theta$, our experiment results are listed in Table~\ref{table:results}.

\begin{table}[hbt]\caption{Experiment results. $\theta$ is the angle of the quoin state; $N_p$ is the total number of $p$-quoins prepared. About half of the prepared $p$-coins are measured in the $X$ basis to prepare the $q$-coin. $q_{\mathrm{th}}$ is the theoretically estimated value based on the estimation of $p$; $q_{\mathrm{exp}}$ is the experimentally estimated value from the obtained $q$-coins; $f_{\mathrm{th}}(p)$ is the theoretically estimated value from the estimation of $p$; $f_{\mathrm{exp}}(p)$ is the experimentally estimated value from the obtained $f(p)$-coins; $N_{f(p)}$ is the number of $f(p)$-coins obtained.}
\begin{tabular}{cccccccc}
  \hline
$\theta$ & $p$ &$N_{p}$& $q_{\mathrm{th}}$&$q_{\mathrm{exp}}$  & $f_{\mathrm{th}}(p)$ & $f_{\mathrm{exp}}(p)$ &$N_{f(p)}$\\
\hline
$0^\circ$&$0.996$&$2.06*10^{7}$&$0.563$&$0.504$&$0.016$&$0.014$&$8.66*{10^5}$\\
$15^\circ$&$0.979$&$1.87*10^{6}$&$0.644$&$0.628$&$0.083$&$0.081$&$8.29*{10^5}$\\
$30^\circ$&$0.929$&$2.07*10^{7}$&$0.756$&$0.746$&$0.262$&$0.257$&$1.05*{10^6}$\\
$45^\circ$&$0.850$&$2.08*10^{7}$&$0.857$&$0.847$&$0.509$&$0.495$&$1.25*{10^6}$\\
$60^\circ$&$0.748$&$2.08*10^{7}$&$0.934$&$0.924$&$0.754$&$0.731$&$1.07*{10^6}$\\
$75^\circ$&$0.630$&$1.99*10^{6}$&$0.983$&$0.974$&$0.933$&$0.901$&$8.90*{10^5}$\\
$90^\circ$&$0.502$&$2.09*10^{7}$&$1.000$&$0.990$&$1.000$&$0.965$&$8.85*{10^5}$\\
$105^\circ$&$0.375$&$2.18*10^{7}$&$0.984$&$0.974$&$0.938$&$0.905$&$9.74*{10^5}$\\
$120^\circ$&$0.258$&$2.08*10^{7}$&$0.938$&$0.926$&$0.766$&$0.737$&$1.06*{10^6}$\\
$135^\circ$&$0.157$&$2.19*10^{7}$&$0.864$&$0.849$&$0.530$&$0.508$&$1.32*{10^6}$\\
$150^\circ$&$0.080$&$2.09*10^{7}$&$0.772$&$0.749$&$0.296$&$0.283$&$1.08*{10^6}$\\
$165^\circ$&$0.033$&$2.08*10^{7}$&$0.678$&$0.633$&$0.126$&$0.120$&$9.45*{10^5}$\\
  \hline
\end{tabular}\label{table:results}
\end{table}

\begin{figure}[bht]
\centering
\resizebox{8cm}{!}{\includegraphics[scale=1]{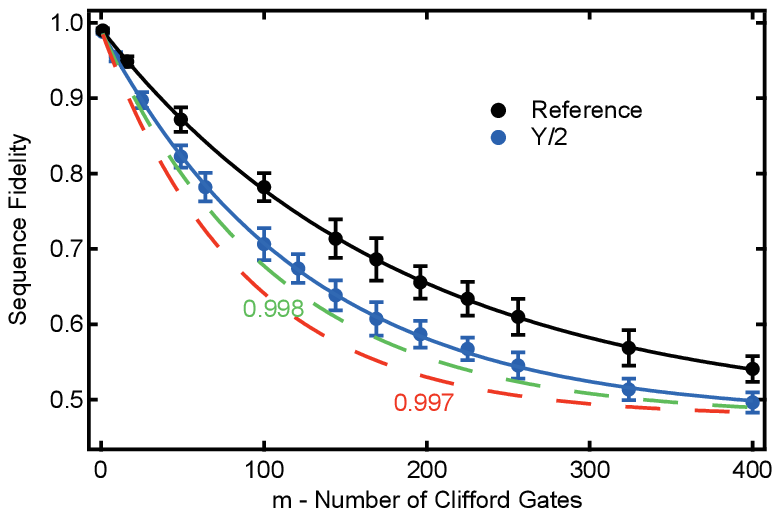}}
\caption{Randomized benchmarking measurement for $R_{\pi/2}^{Y}$ gate fidelity. The reference curve is measured after applying sequences of $m$ random Clifford gates, while the $Y/2$ curve is realized after applying sequences that interleave $R_{\pi/2}^{Y}$ with $m$ random Clifford gates. Each sequence is followed by a recovery Clifford gate in the end right before the final measurement. The number of random sequences of length $m$ in our experiment is $k=100$. Both curves are fitted to $F=Ap^m+B$ with different sequence decay $p$. The data point is the average of the sequence fidelities of the $k=100$ sample sequences, and the error bar shows the standard deviation of the sample. The average single-qubit gate error $r_{s}=r_{ref}/1.875=(1-p_{ref})/2/1.875=0.0014$, and the $R_{\pi/2}^{Y}$ gate error $r_{Y/2}=(1-p_{int}/p_{ref})/2=0.0013$. The dashed lines indicate a gate fidelity of 0.998 and 0.997 respectively.}
\label{fig:RB}
\end{figure}

A randomized benchmarking experiment~\cite{Knill2008,Ryan2009,Magesan2012,BarendsNature} is performed to determine the fidelity of the $\pi/2$ gate around the $Y$ axis, $R_{\pi/2}^{Y}$, which is the most critical gate for the quoin measurement. The randomized gates used in this experiment are chosen from the single-qubit Clifford group. This group contains 24 rotation gates which are composed from rotations around the $X$ and $Y$ axes using the generators: $\{I, +X, +Y, \pm X/2, \pm Y/2\}$. The reference curve is measured after applying sequences of $m$ random Clifford gates, while the $Y/2$ curve is realized after applying sequences that interleave $R_{\pi/2}^{Y}$ with $m$ random Clifford gates. Each sequence is followed by a recovery Clifford gate in the end right before the final measurement. The number of random sequences of length $m$ in our experiment is chosen to be $k=100$. Both curves are fitted to $F=Ap^m+B$ with different sequence decay $p$. The reference decay indicates the average error of the single-qubit gates, while the ratio of the interleaved and reference decay gives the specific gate fidelity. The experiment results are displayed in Fig.~\ref{fig:RB}. The data point is the average of the sequence fidelities of the $k=100$ sample sequences, and the error bar shows the standard deviation of the sample. Each random sequence is measured over 10,000 times to get the sequence fidelity whose error could be neglected. As a result, the average single-qubit gate error $r_{s}=r_{ref}/1.875=(1-p_{ref})/2/1.875=0.0014$, and the $R_{\pi/2}^{Y}$ gate error $r_{Y/2}=(1-p_{int}/p_{ref})/2=0.0013$. The dashed lines indicate a gate fidelity of 0.998 and 0.997 respectively. Therefore, the $R_{\pi/2}^{Y}$ gate fidelity in our experiment is greater than 0.998, and the uncertainty in the gate fidelity is typically 7e-5, determined by bootstrapping.



\bibliography{BibCoin}


\end{document}